\documentclass [12pt,elsart,epsfig]{article}     
\usepackage{epsfig}
\begin{document}
\begin{center}

{\Large \bf Study of $\bar pp \to \eta \eta \pi ^0\pi ^0$ in flight}
\vskip 5mm

{A.V. Anisovich$^c$, C.A. Baker$^a$, C.J. Batty$^a$, D.V. Bugg$^b$, 
 V.A. Nikonov$^c$, A.V. Sarantsev$^c$, V.V. Sarantsev$^c$, 
B.S.~Zou$^{b}$ \footnote{Now at IHEP, Beijing 100039, China} \\
{\normalsize $^a$ \it Rutherford Appleton Laboratory, Chilton, Didcot OX11 0QX,UK}\\
{\normalsize $^b$ \it Queen Mary and Westfield College, London E1\,4NS, UK}\\
{\normalsize $^c$ \it PNPI, Gatchina, St. Petersburg district, 188350, Russia}\\
 }
\end {center}

\begin{abstract}
An analysis of data on $\bar pp \to \eta \eta \pi ^0\pi ^0$
is presented at $\bar p$ beam momenta 600 to 1940 MeV/c.
There is evidence for an $I = 1$, $J^{PC} = 2^{-+}$ resonance in 
$\eta \eta \pi ^0$ with mass
$M = 1880 \pm 20$ MeV and width $255 \pm 45$ MeV, decaying strongly
to $a_2(1320)\eta$; it is too strong to be explained as the high mass tail of 
$\pi _2(1670) \to a_2(1320)\eta$. 
There is tentative evidence also for weak decays to $f_0(1500)\pi$.
It makes a natural partner to the $\eta _2(1860)$.
\end{abstract}

In earlier studies of $\bar pp \to \eta \pi ^0 \pi ^0 \pi ^0$ data in flight,
we have presented evidence for $I = 0$ $J^{PC} = 2^{-+}$ states with
masses 1645, 1860 and 2030 MeV [1,2]. Here, we study the final state
$\eta \eta \pi ^0 \pi ^0$, with the objective of searching for
corresponding $I = 1$ resonances decaying to $\eta \eta \pi ^0$.

The data were taken with the Crystal Barrel detector at LEAR [3].
The analysis techniques run closely parallel to those used in
studying $\eta \pi ^0 \pi^0 \pi ^0$, so we shall refer to
the earlier papers for a description of the experimental set-up and
details concerning amplitude analysis. Here we begin by
outlining the detector briefly.

The $\bar p$ beam interacted in a liquid hydrogen target 4.4 cm long
at the centre of the detector.
The beam was counted by a coincidence between a small proportional
chamber $P$ and a silicon counter $Si_C$ of 5 mm diameter placed 
$\sim 5 $ cm upstream of the target.
Two veto counters 20 cm downstream of the target were used to
select interactions.
The beam intensity was typically $2 \times 10^5 ~\bar p$/s.
A multiwire chamber and a silicon vertex
detector close to the target and covering 98\% of the
solid angle were used for a trigger on neutral final states.
The data-taking rate saturated at $\sim$ 60 events/s.

The essential element of the detector for present purposes was a barrel
of 1380 CsI crystals, each of 16 radiation lengths, detecting photons
with high efficiency down to $<20 $ MeV and with an angular
resolution of $\pm 20$ mrad in both polar and azimuthal angles.
The energy resolution is given by $\Delta E/E = 0.025/E^{1/4}$, where $E$ is
in GeV.
The angular coverage is 98\% of $4\pi$ solid angle.
In order to filter out events which obviously fail to conserve energy, the
total energy in the CsI crystals was summed on-line [4]; a fast trigger
rejected those events with total energy falling $\sim 200$ MeV or more below
that for $\bar pp$ annihilation.

The off-line analysis follows the procedures of Ref. [1], with minor
refinements in data selection based on a Monte Carlo simulation of
cross-talk with background channels.
Events containing exactly 8 photons are selected; the energy of each photon
shower is summed over a block of $3 \times 3$ neighbouring CsI crystals.
Kinematic 7C fits are first made to $\pi ^0 \pi ^0 \eta \gamma \gamma$ and
$\pi ^0 \pi ^0 \pi ^0 \gamma \gamma$, then 8C fits
to $4\pi ^0$, $\eta \pi ^0 \pi ^0 \pi ^0$ and
$\eta \eta \pi ^0 \pi ^0$.
Events fitting $\eta \eta \pi ^0 \pi ^0$ with confidence level $CL > 10\%$
are used.
To eliminate backgrounds from $4\pi ^0$ and $\eta \pi ^0 \pi ^0 \pi ^0$,
events with $CL(3\pi ^0 \gamma \gamma ) > 10^{-3}$ are rejected.
Any surviving events from these channels or from other rare channels are
rejected if they have $CL > CL(\eta \eta \pi ^0 \pi ^0)$.
The Monte Carlo simulation reproduces the observed confidence level
distribution down to 10\%.

\begin{table} [htp]
\begin{center}
\begin{tabular}{cc}
\hline
Source & Background (\%) \\
\hline
$4\pi ^0$ & 0.1  \\
$\eta 3\pi ^0$ & 1.8 \\
$\omega 3\pi ^0$ & 0.7 \\
$\omega \eta 2\pi ^0$ & 6.0 \\
$5\pi ^0$ & 0.7 \\
$\eta 4\pi ^0$ & 4.0 \\
Wrong combinations & 1.7 \\
\hline
\end{tabular}
\caption {Background levels at 1800 MeV/c estimated by the Monte Carlo
simulation using GEANT.}
\end{center}
\end{table}

The normalisation of cross sections is derived from beam counts
$P.Si_C$, the length and density of the target, the observed number
of events, and the Monte Carlo simulation of detection efficiency.
A significant rate-dependence of the number of reconstructed events
is observed, and is consistent with expected pile-up in the CsI
crystals, where scintillations have a long time constant, 100 $\mu s$.
The correction of cross sections for this rate-dependence is 
described at length in Ref. [5].

   Backgrounds are estimated as follows. The Monte Carlo simulation is used
to generate at least 30,000 events in each of 43 exclusive channels 
containing 4--10$\gamma$.
The generated events are fitted both to the original channel (to estimate
detection efficiency) and to all other channels (to estimate cross-talk).
Data are fitted kinematically to all channels having the observed number 
of photons. Numbers of fitted events are then used in a set of 43 $\times$ 
43 linear equations, which fit cross sections for each channel, with 
allowance for backgrounds from other channels;
a constraint is applied that all cross sections are positive or zero.
In practice, the background contributions to these equations are
generally small.

Resulting background levels are illustrated in Table 1 at 1800 MeV/c.
The background comes largely from $\eta 4\pi ^0 $ after the loss
of two photons (4.0\%),
from $\eta \omega \pi ^0 \pi ^0$ after loss of one photon (6\%)
and from $\eta 3\pi ^0$ (1.8\%).
Combinatorics are high in these channels and lead to a background
which follows $\eta \eta \pi ^0 \pi ^0$ phase space closely.
The background is 14\% within errors at all beam momenta;
it is included in the amplitude analysis, but 
results described here are not sensitive to the precise background level.
Numbers of accepted events are shown in Table 2, together with  reconstruction
efficiency $\epsilon$ and cross sections.

\begin{table} [h]
\begin{center}
\begin{tabular}{ccccc}
\hline
Momentum & CM Energy & Events & $\epsilon $ & $\sigma (\eta \eta \pi ^0
\pi ^0)$ \\
(MeV/c) & (MeV) &  & \%  & $(\mu b)$\\
\hline
600  & 1962 & 257  & 10.2 &  $17.3 \pm 2.1$ \\
900  & 2049 & 1996 & 10.1 &  $27.5 \pm 2.4$ \\
1050 & 2098 & 1234  & 10.1 & $24.8 \pm 1.8$ \\
1200 & 2149 & 2387 & 9.8 &   $23.3 \pm 2.0$\\
1350 & 2201 & 1617  & 10.0 & $19.4 \pm 1.7$ \\
1525 & 2263 & 1393  & 9.4 &  $24.1 \pm 1.5$\\
1642 & 2304 & 1669  & 9.5 &  $22.9 \pm 1.4$\\
1800 & 2360 & 1944  & 9.3 &  $19.5 \pm 1.5$\\
1940 & 2409 & 2340  & 8.9 &  $18.6 \pm 1.6$ \\
\hline
\end{tabular}
\caption {Numbers of selected events after
background subtraction; $\epsilon$ is the reconstruction efficiency;
cross sections in the last column have been corrected for backgrounds and
for the $39.3\%$ branching ratio of $\eta \to \gamma \gamma$
and for the branching ratio of $\pi ^0 \to \gamma \gamma$.
Column 2 shows the total energy available in the centre of mass system.}
\end{center}
\end{table}

Mass distributions are shown in Fig. 1 for one beam momentum, 1800
MeV/c.
Full histograms show the final fit and dotted histograms show phase space
distributions.
There is a strong peak at $\sim 1285$ MeV in $\eta \pi \pi$, Fig. 1(a), 
due to either or both of $f_1(1285)$ and $\eta (1295)$.
In $\pi \eta $, there are peaks due to $a_0(980)$ and $a_2(1320)$.
In $\eta \eta$, there is a small enhancement from $f_0(1500)$.
The $\pi \pi$ phase space is limited and this mass distribution
shows little structure; deviations from phase space arise mostly via
reflections from other channels.
The $\pi \eta \eta $ mass distribution shows little structure, but peaks
slightly above phase space at high masses.

\begin{figure}
\vskip -24mm
\centerline{\epsfig{file=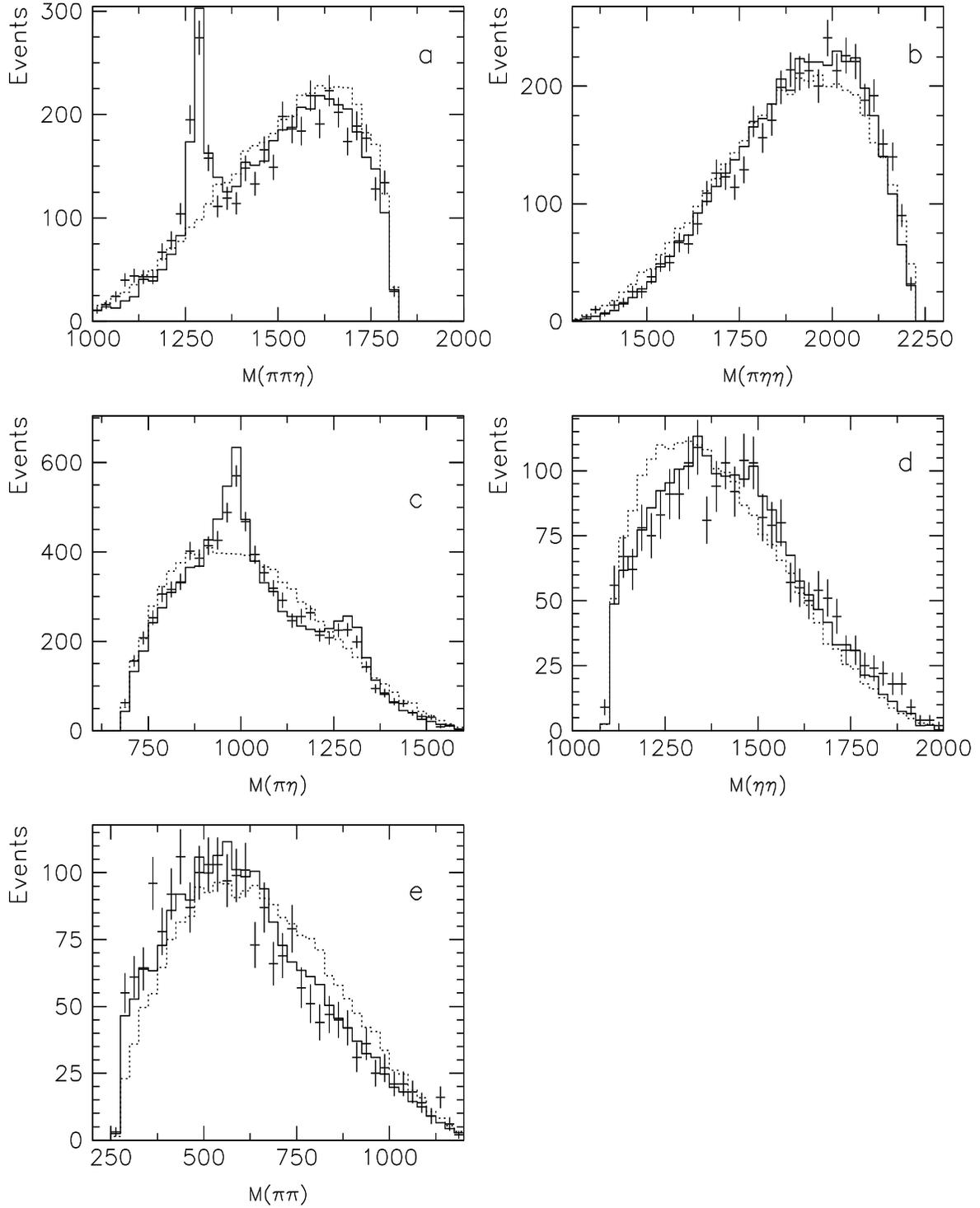,width=16cm}}
\vskip -210mm
\centerline{\epsfig{file=FIG1.EPS,width=16cm}}
\vskip -8mm
\caption{Mass distributions at a beam momentum of 1800 MeV/c for
(a) $ \pi \pi \eta$, (b) $\pi \eta \eta $, (c) $\pi \eta $, (d) $\eta \eta$ and
(e) $\pi \pi$.
Full histograms show the fit and dotted histograms show phase space
distributions. Masses are in MeV.}
\end{figure}

The amplitude analysis fits the following channels to the data:
\begin {eqnarray}
\bar pp &\to& a_0(980)a_0(980) \\
        &\to& a_0(980)a_2(1320) \\
        &\to& f_0(1500)\sigma  ,~f_0(1500) \to \eta \eta \\
        &\to& f_1(1285)\eta   ,~f_1(1285) \to a_0(980)\pi \\
        &\to& \pi (1800)\pi   ,~\pi (1800) \to f_0(1500)\eta \\
        &\to& \pi (1800)\pi   ,~\pi (1800) \to a_0(980)\eta \\
        &\to& \pi _2(1880)\pi ,~\pi _2(1880) \to a_2(1320)\eta \\
        &\to& \pi _2(1880)\pi ,~\pi _2(1880) \to [f_0(1500)\pi ]_{L = 2} \\
        &\to& X (2200)\pi    ,~X \to a_2(1320)\eta .
\end {eqnarray}
Masses and widths of established states are taken from the Particle Data Group
(PDG) [6].
The peaks fitted to $f_1(1285)$, $a_0(980)$ and $a_2(1320)$ in
Figs. 1(a) and (c) slightly over-estimate the data.
Inclusion of mass resolution improves the fit to $f_1(1285)$ marginally
but has negligible effect for $a_0(980)$ and $a_2(1320)$.
Altering the widths of the latter two states also has negligible
effect; although the fits to peaks in Fig. 1(a) and (c) may be improved,
the fit to other features gets correspondinly worse.
Likewise, altering decay branching ratios in equations (6) and (7)
can improve the fit at one momentum, but the overall fit is made
to all momenta simultaneously.

In reaction (3), $\sigma$ is a shorthand for the broad $f_0(400-1200)$ of
the PDG; it has been fitted with the
parametrisation of Zou and Bugg [7].
Further final states have been tried, but are not required.
As an example, $\eta (1295)\eta$ has been tried, replacing or supplementing
$f_1(1285)\eta$; also decays of both $\eta (1295)$ and $f_1(1285)$ to
$\sigma \eta$ have been tried. There is no significant change to the fit,
but we cannot exclude some small component due to $\eta (1295)\eta$.
The background underneath the $f_1(1285)$/$\eta (1295)$ is too large
to allow the possibility of fitting this channel in terms of partial waves in
the production process.
Contributions have been tried from $\pi _2(1670)$ decaying to $a_0(980)\eta$,
$f_2(1270)\pi$, $f_0(1370)\pi$ and $f_2(1565)\pi$, but are not needed.
In the last three cases, this is hardly surprising; $f_2(1270)$ has
a very small decay branching ratio to $\eta \eta$, and decays of $f_0(1370)$
and $f_2(1565)$ to two-body channels are believed to be weak.
The known decay $\pi (1800) \to f_0(1370)\pi$, $f_0(1370) \to \eta \eta$ 
was also tried, but found to be negligible.

There are not sufficient events for a full partial wave decomposition of
both production and decay.
As an approximation, the production process has been ignored
and only the decays of resonances have been fitted.
The objective is to try to identify characteristic decay signatures
from observed angular distributions.
Taking reaction (7) as an example, it is assumed that the
$\pi _2(1880)$ is produced through three helicity amplitudes, with
components of spin along the beam direction 0, $\pm 1$ and $\pm 2$.
The decay angular distribution of each component is fitted in full,
using the method of Wick rotations described in Ref. [1].
In essence, this involves (i) a rotation of axes from the beam
direction to that of the resonance in the overall centre of
mass frame, (ii) a Lorentz boost to the rest frame of the resonance,
(iii) a rotation back through precisely the same angles as are used
in step (i). In the Lorentz boost, helicity amplitudes are unchanged;
in steps (i) and (iii), there is a cancellation of rotation matrices.
As a result, decay amplitudes may be expressed simply in terms of
Clebsch-Gordan coefficients. The Wick rotation essentially takes care of
the Lorentz transformation from the centre of mass to the rest frame of
the resonance.
A second Wick rotation is used for the subsequent decay of $a_2(1320)$ to
$\eta \pi$.

Because beam and target are unpolarised, there is no dependence of
cross sections on the azimuthal angle around the beam.
Consequently, it may be shown that there are no interferences between
different helicities along the beam direction in the final state.
Interferences between channels for a given helicity are allowed.
However, these occur only over those limited parts of phase space where
both amplitudes are large; in practice, most of the
interferences have negligible effect and are dropped.
There is, however a large interference between the $a_0(980)a_2(1320)$
channel and $\pi _2(1880) \to a_2(1320)\eta$.
This interference is phase sensitive and helps determine the mass and width 
of $\pi _2(1880)$ precisely.
For a pair of channels such as (5) and (6), where a single resonance
decays through two different modes, they must be fully coherent
and this coherence is retained in the fit.
Every channel is fitted with a coupling constant at each beam momentum,
and a phase for those cases where interferences survive.

\begin{table} [htp]
\begin{center}
\begin{tabular}{ccccccccc}
\hline
Channel & 900 & 1050 & 1200 & 1350 & 1525 & 1642 & 1800 & 1940 \\\hline
 $a_0(980)a_0(980)$  & 10  &   8 & 12 & 40 & 23 & 40 & 44 & 38\\
 $a_0(980)a_2(1320)$ & -   &   - & -  & 7  & 17 & 33 & 89 & 158\\
 $f_0(1500)\sigma$   &   8 &   4 &  6 & 15 & 17 & 24 & 11 & 14\\
 $f_1(1285)\eta$     & 153 & 113 & 173&177 &149 &264 &305 &300 \\
 $\pi (1800) \to f_0(1500)\eta$ &   0 &   4 &   8&  2 &  3 &  1 &  2 &  6\\
 $\pi (1800) \to a_0(980)\eta$  &  25 &  11 &  21&  2 &  7 &  9 &  1 &  9\\
 $\pi _2(1880) \to a_2(1320)\eta$& 169 & 152 & 246& 51 & 28 & 38 & 56 &  0 \\
 $\pi _2(1880) \to f_0(1500)\pi$ &   3 &   9 &  16&  4 &  2 & 10 &  3 &  8 \\
 $X (2200)$          & -   & -   & -  & -  &12 &25 &31 &40 \\\hline
\end{tabular}
\caption {Changes in log likelihood when each channel is dropped
from the fit and remaining contributions are re-optimised; the top line of
the table indicates beam momenta in MeV/c.}
\end{center}
\end{table}
We now turn to the detailed features of the data.
Initial fits were made with channels (1)--(6). 
An additional contribution from $\sigma \sigma$ final states was tried, 
with one $\sigma \to \pi ^0 \pi ^0$ and the second $\sigma \to \eta \eta$;
this had no significant effect.
The contribution from $\pi (1800)$ is small, but just significant.
For example, it improves log likelihood by 29 at 1200 MeV/c
in the final fit reported below (a $6.5\sigma$ effect statistically for
3 fitted parameters).
Table 3 illustrates significance levels by giving changes in log
likelihood when each channel is dropped in turn from the final fit
and all other components are re-optimised.
Our definition of log likelihood is such that it changes by 0.5 for
a one standard deviation change in one fitted parameter.

\begin{figure} 
\vskip -12mm
\centerline{\epsfig{file=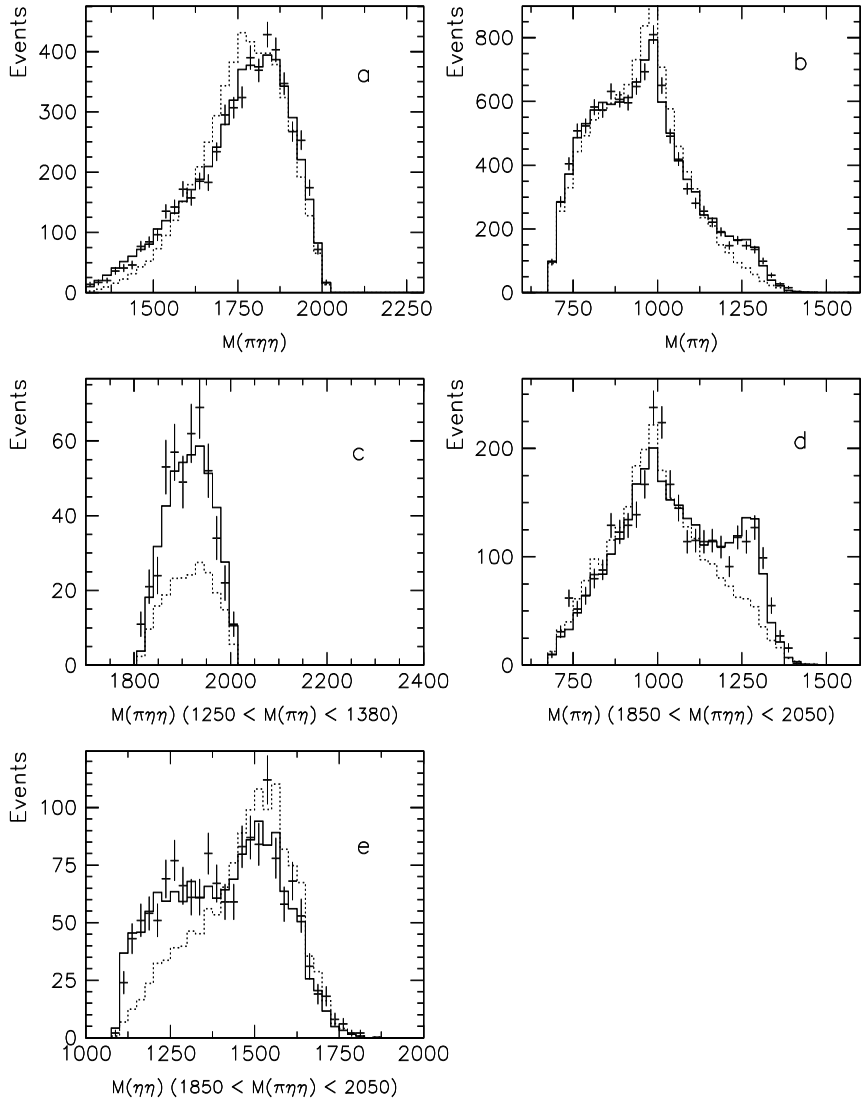,width=12cm}}
\vskip -157.75mm
\centerline{\epsfig{file=FIG2.EPS,width=12cm}}
\vskip -11mm

\caption{Full histograms show fits at 1200 MeV/c compared with data;
dotted histograms show fits without $\pi _2(1880)$;
(a) the $\pi \eta \eta $ mass distribution, (b) the $\pi \eta$ mass
distribution,
(c) the $\pi \eta \eta  $ mass distribution for events with a
$\pi \eta $ combination within $\pm 65$ MeV of $a_2(1320)$,
(d) and (e) $\pi \eta$ and $\eta \eta$ mass distributions for events with
$M (\pi \eta \eta )$ 1850-2050 MeV. Units of mass are MeV.}
\end{figure}

At this stage, it became obvious at beam momenta 900--1200 MeV/c that
several features of the data are fitted poorly, particularly the
$\pi \eta \eta $ mass distribution. This is illustrated in Fig. 2
with several distributions at 1200 MeV/c. In each case, the absolute
normalisation of histograms is taken from the fit.
There is a clear requirement for something with
$\pi \eta \eta $ mass around 1880 MeV with decays to $a_2(1320)\eta$.
It improves log likelihood by 246 at 1200 MeV/c and by similar
large amounts at 900 and 1050 MeV/c.
Its mass and width are best determined by the data at 1050 and
1200 MeV/c;
at 900 MeV/c, the upper side of the resonance is cut off.
A further marginal improvement is obtained by adding channel (8),
$\pi _2(1880) \to [f_0(1500)\pi]_{L = 2}$.
The possible decay $\pi _2(1880) \to f_0(1370)\pi$, $f_0(1370) \to \eta \eta$
was tried, but found negligible; in view of the known weak decay of $f_0(1370)$
to $\eta \eta$, this is not surprising.

\begin{figure}
\vskip -10mm
\centerline{\epsfig{file=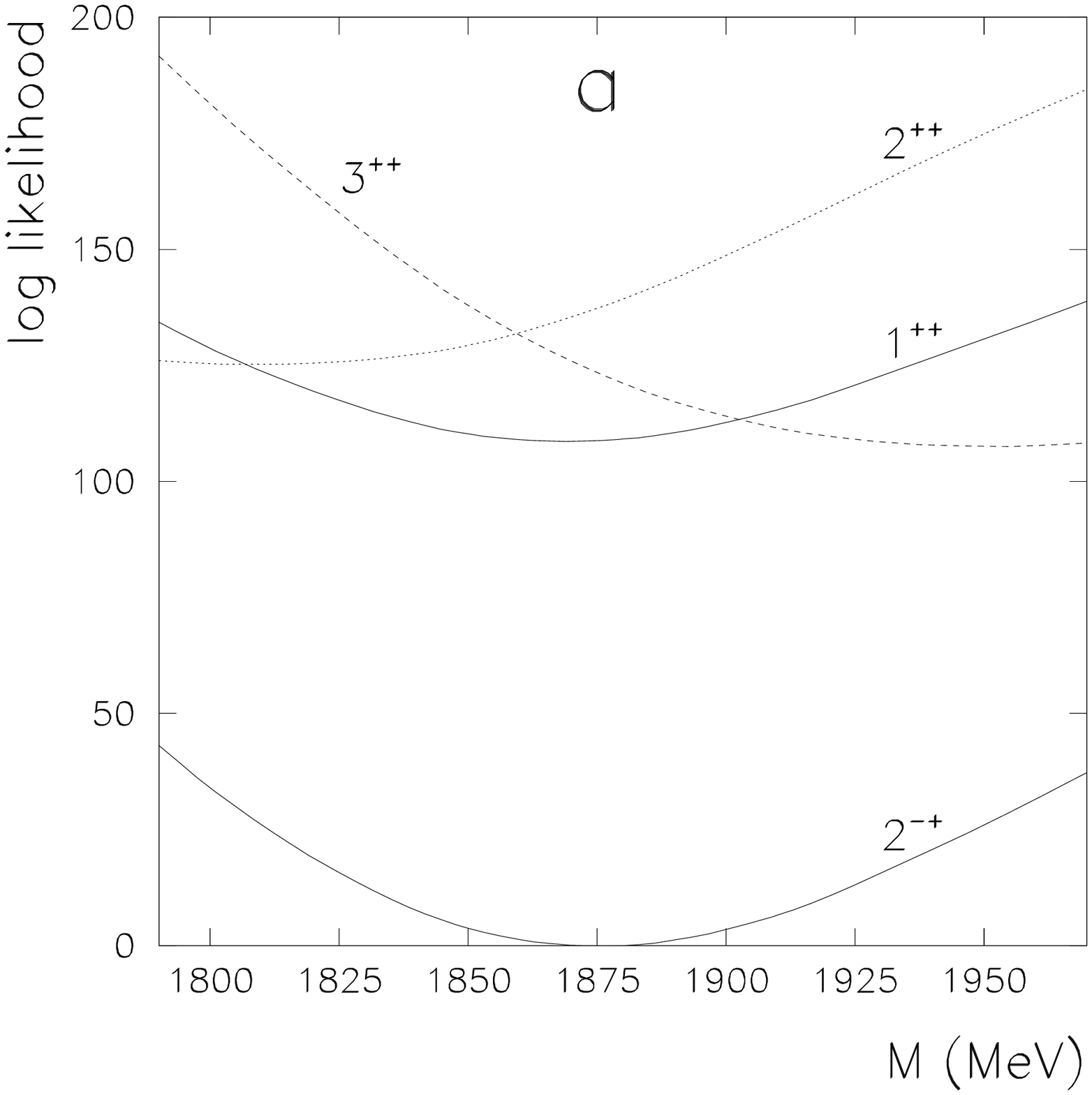,height=7cm}\hspace{0.5cm}
            \epsfig{file=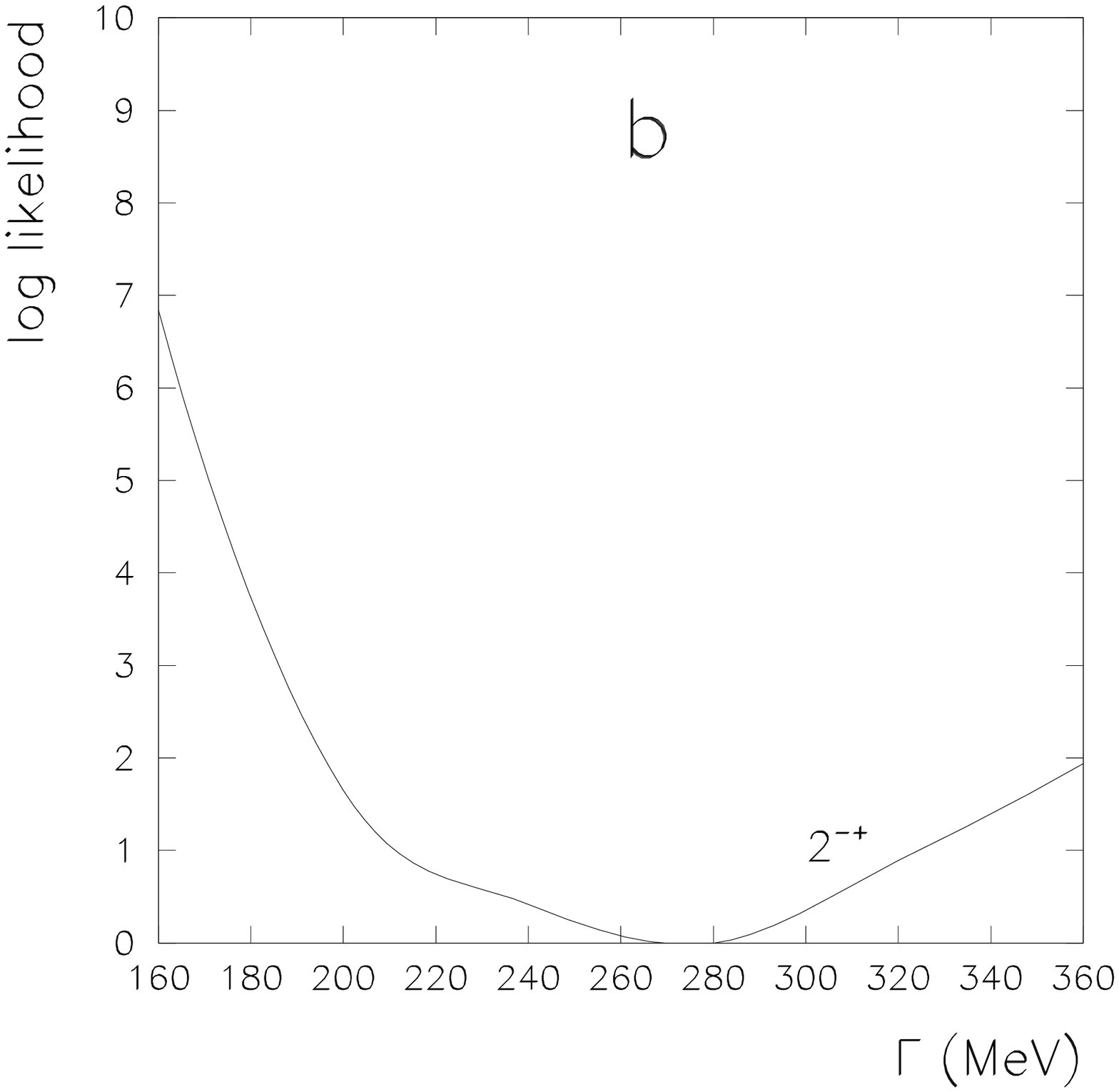,height=7cm}}
\vskip -70.35mm
\centerline{\epsfig{file=FIG3A.EPS,height=7cm}\hspace{0.5cm}
            \epsfig{file=FIG3B.EPS,height=7cm}}
\vskip -8mm
\caption{(a) Variations of log likelihood with mass, summed over data at 1050
and 1200 MeV/c, for quantum numbers $J^{PC} = 2^{-+}$, $1^{++}$, $2^{++}$
or $3^{++}$; (b) variation of log likelihood with width for $J^{PC} = 2^{-+}$.}
\end{figure}

The mass of the fitted $a_2(1320)\eta$ signal  is so close to
threshold that only decays with orbital angular momentum
$L = 0$ are likely, i.e. quantum numbers
$J^{PC} = 2^{-+}$.
For $L = 1 $ decays, alternative quantum numbers are
$1^{++}$, $2^{++}$ and $3^{++}$.
Fig. 3(a) shows the variation of log likelihood with mass, summed over
data at 1050 and 1200 MeV/c,  for these quantum numbers; the vertical scale 
is adjusted to zero at the optimum for $2^{-+}$.
There is a well defined optimum for these quantum numbers, giving
$M = 1880 \pm 16$ MeV, $\Gamma = 255 \pm 45$ MeV.  
The errors cover
the variations over the four beam momenta 900-1350 MeV/c.
For other quantum numbers, curves of Fig. 3 use the optimum widths,
including appropriate centrifugal barrier factors.
Log likelihood is worse by
at least 109 for $1^{++}$, 125 for $2^{++}$ and 112 for $3^{++}$;
each of these differences from $2^{-+}$  corresponds  to at least a 14
standard deviation effect.
For $L = 1$ decays, it is the centrifugal barrier which prevents
an adequate fit to the  $a_2(1320)\eta$ threshold region.
A second point is that $1^{++}$ and $3^{++}$ states cannot be produced
with orbital angular momentum $\ell = 0$ in the production process, since
this would require exotic quantum numbers $1^{-+}$ and $3^{-+}$ for the initial
state and these are
forbidden for $\bar pp$;  a centrifugal barrier of at least $\ell = 1$ is
required in the production process and plays a significant role in ruling
out these quantum numbers.

Fig. 3(b) shows the variation of log likelihood with the width fitted to
$\pi _2(1880)$. Our experience elsewhere is that $1/\Gamma$ follows an
approximately normal distribution; we use this result in determining
an estimate of the optimum width, $\Gamma = 255 \pm 45 $ MeV.
The error is mostly statistical, but includes a small systematic component.

The VES collaboration has also reported a threshold enhancement in
$a_2(1320)\eta$ with $J^P = 2^-$ in their $\eta \eta \pi ^-$ data [8].
An even clearer threshold peak is observed in their $\eta \pi ^+ \pi ^- \pi ^0$
data, where $a_2(1320) \to \pi ^- \pi ^- \pi ^0$; there is also
a strong peak at the same mass in the $f_2(1270)\pi$ D-wave in their $4\pi$
data [9].
Daum et al. [10] also observed a peak at 1850 MeV in the $f_2(1270)\pi$
D-wave with a width of $\sim 240$ MeV. 
They interpreted it as arising from interference between $\pi _2(1670)$
and a higher $\pi _2(2100)$.
That interpretation fails to fit the present data, since the peak at 1880
MeV in $a_2(1320)\eta$ cannot be explained as the high mass tail of
$\pi _2(1670)$, as we show below.
Furthermore, the peak in the $f_2(1270)\pi$ D-wave cannot plausibly be
explained as due only to the high mass tail of $\pi _2(1670)$, 
since our calculations show that the $L = 2$ centrifugal barrier is not 
strong enough to shift the peak position by $\sim 180$ MeV.

\begin{figure}
\vskip -10mm

\centerline{\epsfig{file=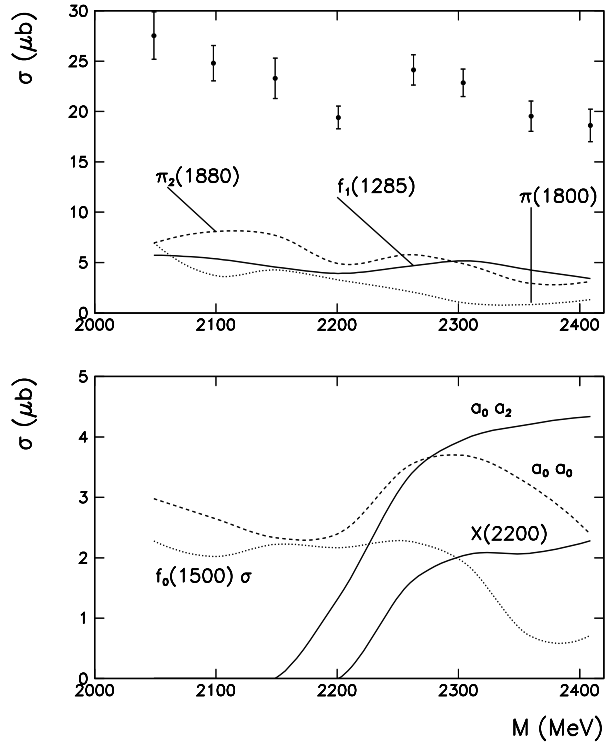,height=11cm}}
\vskip -110.3mm
\centerline{\epsfig{file=FIG4.EPS,height=11cm}}
\vskip -12mm
\caption{Cross sections fitted to individual channels; points with errors
show the overall cross section for $\eta \eta \pi ^0 \pi ^0$.}
\end{figure}

Fitted cross sections are illustrated in Fig. 4.
There is a  strong contribution from $\pi _2(1880)$ at low beam momenta,
falling at high momenta.  The earlier analysis of $\eta \pi ^0 \pi ^0 \pi ^0$
found an analogous peak due to $\eta _2(1860)$
at beam momenta 900--1200 MeV/c [2].

In initial fits to data, the decay branching ratio of $\pi _2(1880)$
between $a_2(1320)\eta$ and $f_0(1500)\pi$ was fitted freely for
beam momenta 900--1350 MeV/c. The weighted average was then formed and
the final fit is made with this weighted mean.
The result, corrected for all charge states and  the branching
ratio of $a_2(1320) \to \eta \pi$,   is
\begin {equation}
BR[\pi _2(1880) \to f_0(1500)\pi ]/
BR[\pi _2(1880) \to a_2(1320)\eta] = 0.28 ^{+0.20} _{-0.15}.
\end {equation}
The significance level of the decay to $f_0(1500)\pi$ is $\sim 3\sigma$ because
the errors are not symmetric about the mean.
If this decay is omitted, other components of the fit change very little.

The branching ratio of $\pi (1800)$ between $f_0(1500)\pi$ and $a_0(980)\eta$
has been treated likewise, with the result
\begin {equation}
BR[\pi (1800) \to a_0(980)\eta]/
BR[\pi (1800) \to f_0(1500)\pi ] = 0.030 ^{+0.014}_{-0.011}.
\end {equation}
The numerical value of this ratio is small because the $f_0(1500)$ couples
only weakly to $\eta \eta$.
Our value is somewhat smaller than the value quoted by VES [8], namely
$0.08 \pm 0.03$, but the errors of both determinations are sizeable.
The mass and width of the $\pi (1800)$
are not determined accurately from the present data, and are
therefore set to PDG values, namely M = 1801 MeV, $\Gamma = 210$ MeV.

\begin{figure}
\vskip -10mm
\centerline{\epsfig{file=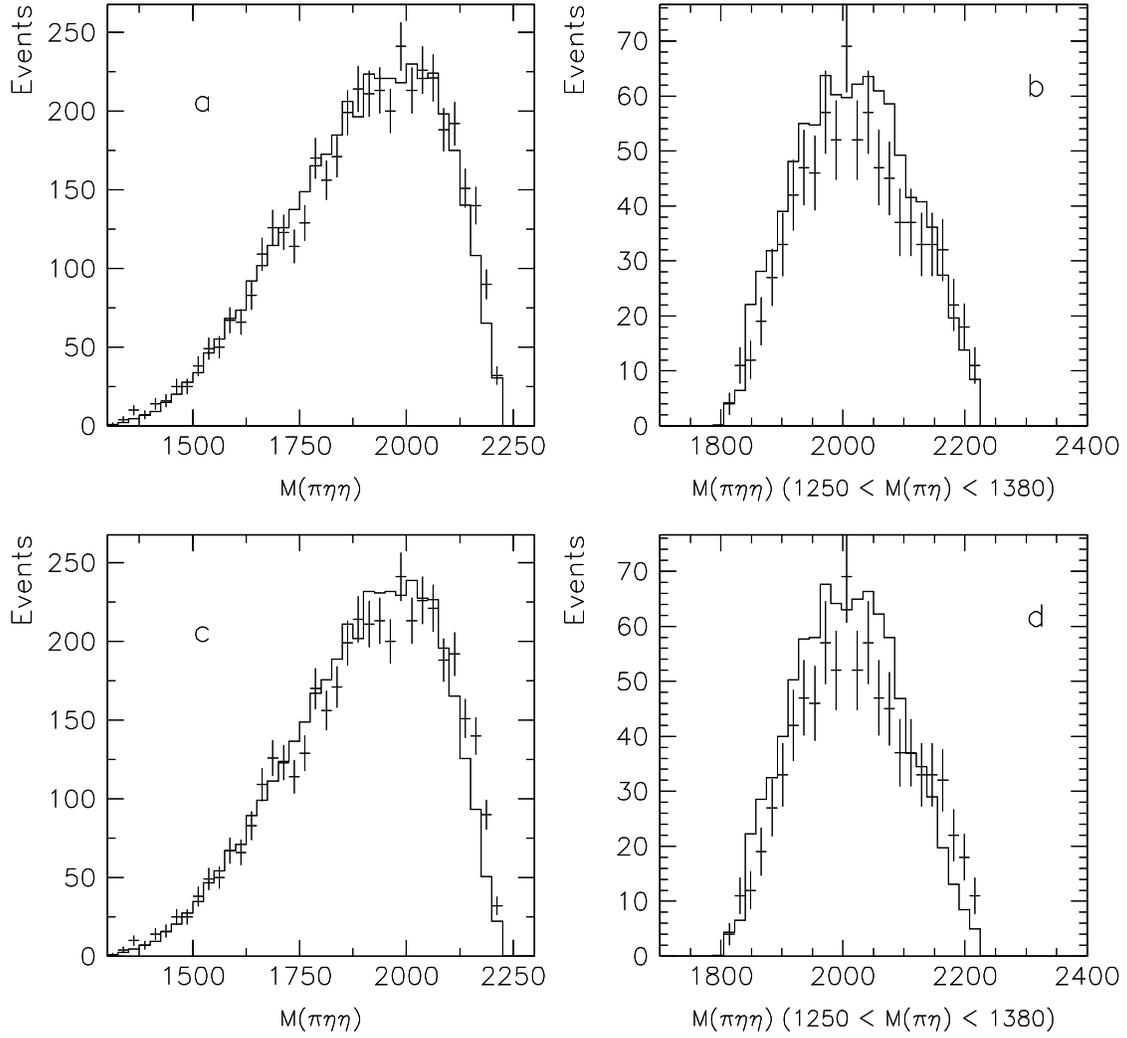,height=15.2cm}}
\vskip -152.3mm
\centerline{\epsfig{file=FIG5.EPS,height=15.2cm}}
\vskip -10mm
\caption{Histograms show fits at 1800 MeV/c  compared with data,
(a) and (b) including channel (9), (c) and (d) without it;
(a) and (c) show the  $\pi \eta \eta $ mass distribution,
(b) and (d) show this mass distribution for events with a
$\pi \eta $ combination within $\pm 65$ MeV of $a_2(1320)$.
Masses are in MeV. }
\end{figure}
We turn now to the higher beam momenta, 1525--1940 MeV/c.
There, the $a_0(980)a_2(1320)$
channel makes a highly significant contribution.
However, the fit to the $a_2(1320)$ peak is not quite perfect without
some further small contribution from channel (9), i.e. something of mass
2050-2200 MeV decaying to $a_2(1320)\eta$.
This is a mass region where many resonances are expected with $J^P$ up
to $4^+$.
At all beam momenta 1525-1940 MeV, the fitted mass for this extra component
optimises close to the top of the available mass range.
Taking these four momenta
together, the optimum is at a mass $M  = 2200 \pm 40$ MeV with width
$\Gamma = 225 \pm 50$ MeV.
Fig. 5 illustrates the small improvement in the fit at a beam momentum of
1800 MeV/c from adding channel (9).
As one sees from Figs. 5(a) and (b), the fit including channel (9)
makes some improvement, but does not succeed in removing the discrepancy
completely, possibly indicating the need for more than one contribution, 
presently unresolved; a better fit may be improved by increasing the mass of
channel (9) above 2200 MeV, but appears unphysical.

Fig. 6 show the variation of log likelihood with the mass fitted
to channel (9) for quantum numbers $2^{-+}$, $1^{++}$, $2^{++}$ and
$3^{++}$. For all $J^{PC}$, the improvement in the fit is small, but
significant.
We find that the quantum numbers giving the best fit are $J^P = 1^+$ or
$2^+$ decaying to $a_2(1320)\eta$ with $L = 1$.
Unfortunately, the distinction between these two possibilities
is poor. Summing over data at the four momenta 1525-1940 MeV/c,
log likelihood is better by 7 for $J^P = 1^+$, but this is a barely
significant difference; at two momenta $1^{++}$ gives slightly the
better fit and at the other two momenta $2^{++}$ is preferred.
Data for $I = 0$ channels [11,12] have located an $f_2$ resonance at 2240 MeV
and a $2^{-+}$ resonance at 2267 MeV.
Channel (9) could be due to the $I = 1$ analogues of either of these states.
\begin{figure}
\vskip -10mm
\centerline{\epsfig{file=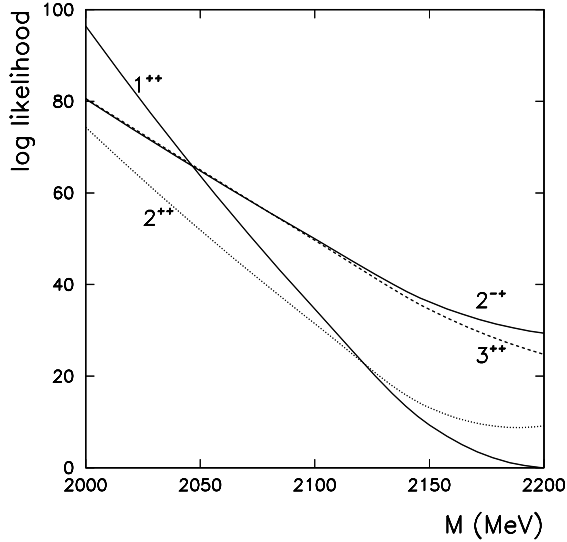,height=8cm}}
\vskip -80.3mm
\centerline{\epsfig{file=FIG6.EPS,height=8cm}}
\vskip -6mm
\caption{Variations of log likelihood with mass, summed over data from 1525
to 1940 MeV/c, for quantum numbers $J^{PC} = 2^{-+}$, $1^{++}$, $2^{++}$
or $3^{++}$.}
\end{figure}

Data on $\bar pp \to \eta 3\pi ^0$ [2] at 1525-1940 MeV/c
show features similar to the present data.
There is an analogous high mass contribution in $\eta \pi \pi$
which peaks at the highest available mass at all momenta.
The statistics of those data are higher by a factor 10,
and it is possible to identify the preferred quantum numbers of the high
mass contribution as $2^{++} \to f_2(1270)\eta$ and $a_2(1320)\pi ^0$.
This is a hint that $2^{++}$ may contribute to present data.
Final fits to present data are therefore made with this $J^{PC}$,
but are almost indistinguishable from $1^{++}$.
We see the possibility that the contribution from the 2050--2200 mass range
could come from more than one state.

An obvious question is whether a wider  $\pi _2$ alone could fit
data at all momenta.
This is not the case.
Data at 1642--1940 MeV/c can be fitted with a single broad resonance
with mass $\sim 2150$ MeV and width 500 MeV. However, the data at 900--1200
MeV/c cannot be fitted with the same broad resonance.
If this attempted, log likelihood gets worse by typically 25 at each momentum.
The fit to the $\pi \eta \eta $ mass distribution then lies very close to phase
space and the peaks of Fig. 2(a) and (c) are not reproduced.
Another point is that data at 1525-1940 MeV/c prefer a
contribution in the $\pi \eta \eta $ mass range 2050--2300 MeV
with $J^{PC} = 2^{++}$ or $1^{++}$ rather than $2^{-+}$.

A second question is whether the signal fitted as $\pi _2(1880)$ could
be due to the high mass tail of $\pi _2(1670)$ and the opening of the
$a_2(1320)\eta$ channel. This possibility has been tried but fails.
The effect of the extra channel is to add to the Breit-Wigner denominator
of the $\pi _2(1670)$ a width $\Gamma (a_2\eta )$;
assuming ideal mixing between $f_2(1270)$ and $a_2(1320)$,
this width is related to the decay width to $f_2\pi$ by their relative phase space:
\begin {equation}
\Gamma (a_2\eta ) = (0.8)^2(k_{\eta }/k_{\pi })\Gamma (f_2\pi );
\end {equation}
here values of k are average momenta for the two decay channels in the
resonance rest frame, folding in the
line-shapes of $f_2(1270)$ and $a_2(1320)$.
The factor $(0.8)^2$ allows for the non-strange content of the $\eta$.
The value of $k_{\eta}/k_{\pi}$ 
rises rapidly from threshold and is 0.35 at 1880 MeV.
However, the line-shape of $\pi _2(1670)$ suppresses high momenta.
The mean value of $k_{\eta}/k_{\pi}$, averaged over the line-shape of $\pi
_2(1670)$, is 0.22.
So  $\Gamma (a_2\eta )$ is small and can be neglected to a good approximation
in the Breit-Wigner amplitude for $\pi _2(1670)$.
If the data are fitted with the $\pi _2(1670)$ (using the PDG
mass and width) instead of $\pi _2(1880)$, there is an unacceptable increase in
log likelihood: 108 at 1050 MeV/c and 97 at 1200 MeV/c.
The $\pi _2(1670)$ is too narrow to reproduce the high mass
side of the $\pi _2(1880)$.
When $\Gamma (a_2\eta )$ is included in the denominator of the Breit-Wigner
amplitude for the $\pi _2(1670)$, it suppresses the amplitude above the
$a_2\eta$ threshold, making the resonance appear narrower and making
the discrepancy with present data even worse.

The explanation of $\pi _2(1880)$ as the high mass tail of $\pi _2(1670)$ is
also ruled out by the large observed cross section for production of 
$\pi _2(1880)$.
We have earlier studied $\bar pp \to 4\pi ^0$; results at one beam
momentum were presented in Ref. [13]. 
In those data, there is a well defined signal due to 
$\bar pp \to \pi _2(1670)\pi ^0$, $\pi _2(1670) \to f_2(1270)\pi ^0$.
It contributes a fraction 15--20\% of the $4\pi ^0$ data at all beam
momenta and a cross section $\sim 20 \mu b$.
The cross section for  $\bar pp \to \pi _2(1670)\pi ^0$, $\pi _2(1670) \to
a_2(1320)\eta$ may then be estimated using equn. (12).
It is necessary to allow for the 14.5\% branching ratio of $a_2(1320) \to \eta
\pi $ and the 28.2\% branching ratio of $f_2(1270) \to \pi ^0 \pi ^0$.
Folding in the line-shape of $\pi _2(1670)$, the predicted cross section
in present data is $\sim 0.7 \mu b$. 
This is a factor 5--10 lower than the observed cross section shown in Fig.
4(a).
In our earlier work on $\bar pp \to \eta 3\pi ^0$ , the same argument ruled 
out the explanation of $\eta _2(1860)$ as the high mass tail of 
$\eta _2(1645)$ [1];
there, the $\eta _2(1860) \to f_2(1270)\eta$ signal is likewise a factor 
$\sim 11-22$ larger than that predicted for $\eta _2(1645) \to f_2(1270)\eta$.
  
We have tried fitting present data with a Flatt\' e formula for
$\pi _2(1880)$ where $\Gamma (a_2\eta )$ is related to $\Gamma (f_2\pi )$
by equn. (12). 
This increases the fitted mass by 5 MeV, because of the suppression of the
upper side of the resonance by the rising phase space for $a_2\eta$.
However, the magnitude of possible contributions of the $f_2(1270)\pi$ D-wave
and other channels are not known and could have larger effects in the
Flatt\' e formula than $a_2\eta$. Hence we simply include the possible 
5 MeV shift into an overall error of $\pm 20$ MeV for the mass.

The observed strong decay of $\pi _2(1880)$ to $a_2(1320)\eta$
makes it a natural partner for $\eta _2(1860)$, which decays strongly
to $f_2(1270)\eta$ and $a_2(1320)\pi$ [2].
This makes it unlikely that the $\eta _2(1860)$ is the nonet (dominantly
$s\bar s)$  partner of $\eta _2(1645)$.
Our earlier publication, Ref. [2], finds three $\eta _2$ resonances
at 1645, 1860 and 2030 MeV.
There are too many $2^{-+}$ states in a
narrow mass range for all to be accomodated as $q\bar q$ states.
We conjectured in Ref. [2] that $\eta _2(1860)$ is a candidate for a
hybrid expected at roughly this mass. Decays $\eta _2(1860) \to
f_2(1270)\eta$ and $a_2(1320)\pi$ and decays $\pi _2(1880) \to a_2(1320)\eta$
are favoured for a hybrid in the flux-tube model [14].

In summary, we find evidence for a new state having $I = 1$, $J^{PC} = 2^{-+}$,
with $M = 1880 \pm 20$ MeV, $\Gamma = 255 \pm 45$ MeV, decaying strongly
to $a_2(1320)\eta$; there is also a possible weak decay mode to
$f_0(1500)\pi$, but this is only a $3\sigma$ effect.
There is some evidence for a further contribution with mass $\sim 2200$ 
decaying to $a_2(1320)\eta$, but it is weak and we are unable to 
distinguish clearly between quantum numbers $1^{++}$ and $2^{++}$.

We thank the Crystal Barrel Collaboration for
allowing use of the data.
We acknowledge financial support from the British Particle Physics and
Astronomy Research Council (PPARC).
We wish to thank Prof. V. V. Anisovich for helpful discussions.
The St. Petersburg group wishes to acknowledge financial support from PPARC and
INTAS grant RFBR 95-0267.


\begin {thebibliography}{99}
\bibitem {1} J. Adomeit et al., Zeit. Phys. C 71 (1996) 227.
\bibitem {2} A.V. Anisovich et al., Phys. Lett. B 477 (2000) 19.
\bibitem {3} E. Aker et al., Nucl. Instr. A 321 (1992) 69.
\bibitem {4} C.A. Baker, N.P . Hessey, C.N. Pinder and C.J. Batty, Nucl. Instr.
A394 (1997) 180.
\bibitem {5} A.V. Anisovich et al., Nucl. Phys. A662 (2000) 344.
\bibitem {6} Particle Data Group, Euro. Phys. Journ. C15 (2000) 1.
\bibitem {7} B.S. Zou and D.V. Bugg, Phys. Rev. D 48 (1993) R3948.
\bibitem {8} D. Amelin et al., Phys. At. Nucl. 59 (1996) 976.
\bibitem {9} D. Ryabchikov, Ph.D. thesis, IHEP, Protvino (1999).
\bibitem {10} C. Daum et al., Nucl. Phys. B182 (1981) 269.
\bibitem {11} A.V. Anisovich et al., Phys. Lett. B491 (2000) 47.
\bibitem {12} A.V. Anisovich et al., Phys. Lett. B491 (2000) 40.
\bibitem {13} C.A. Baker et al., Phys. Lett. B 449 (1999) 114.
\bibitem {14} P.R. Page, E.S. Swanson and A.P. Szczepaniak, Phys. Rev. D 59
(1999) 034016.
\end {thebibliography}
\end{document}